\documentclass[journal]{IEEEtran}
\ifCLASSINFOpdf
\else
\fi
\usepackage{bm}
\usepackage{epstopdf}
\usepackage{multirow}
\usepackage{cite}
\usepackage{url}
\usepackage{stfloats}
\usepackage{color}
\usepackage{epsfig}
\usepackage{graphicx}
\usepackage{subfigure}
\usepackage{psfig}
\usepackage{epsf}
\usepackage{slashbox}
\usepackage{float}
\usepackage{amssymb }
\usepackage{algorithm}
\usepackage{algorithmic}
\usepackage[cmex10]{amsmath}
\usepackage[justification=centering]{caption}
\begin{document}
\title{{Computation Bits Maximization in a Backscatter Assisted Wirelessly Powered MEC Network }}
\author{\IEEEauthorblockN{Liqin Shi, Yinghui Ye, Xiaoli Chu, and Guangyue Lu} \vspace{-15pt}
\thanks{Liqin Shi, Yinghui Ye, and Guangyue Lu  are with the Shaanxi Key Laboratory of Information Communication Network and Security, Xi'an University of Posts \& Telecommunications, China. Xiaoli Chu is with the Department of Electronic and Electrical Engineering,
The University of Sheffield, U.K. The corresponding author is Yinghui Ye (connectyyh@126.com). }
}
\markboth{ }
{Ye\MakeLowercase{\textit{et al.}}: Computation Bits Maximization in NOMA based WPT-MEC Networks}
\maketitle

\begin{abstract}
In this paper, we introduce a backscatter assisted wirelessly powered mobile edge computing (MEC) network, where each edge user (EU) can offload task bits to the MEC server via hybrid harvest-then-transmit (HTT) and backscatter communications. 
In particular, considering a practical non-linear energy harvesting (EH) model and a partial offloading scheme at each EU,
we propose a scheme to maximize the weighted sum computation bits of all the EUs by jointly optimizing the backscatter reflection coefficient and time, active transmission power and time, local computing frequency and execution time of each EU. {\color{black}By introducing a series of auxiliary variables and using the properties of the non-linear EH model, we transform the original non-convex problem into a convex one and derive closed-form expressions for parts of the optimal solutions.} Simulation results demonstrate the advantage
of the proposed scheme over benchmark schemes in terms of weighted sum computation bits.
\end{abstract}
\vspace{-10pt}
\IEEEpeerreviewmaketitle
\vspace{-5pt}
\section{Introduction}
\IEEEPARstart{W}{irelessly} powered mobile edge computing (MEC) has been deemed  an emerging technology for the Internet of Things (IoT), since it can provide energy  and enhance  computation capacity for IoT devices. {\color{black}In a wireless powered MEC network}, the IoT devices (also referred to as edge users (EUs)) harvest energy from the  energy source (i.e., power beacon (PB)) and then utilize the harvested energy to execute their task locally and/or offload their task bits to an MEC server so that the task can be executed within the given budget while not consuming their  battery power \cite{7442079}. 

{\color{black}To date, several works have studied the resource allocation in wirelessly powered MEC networks \cite{7442079,8771176,8334188,8234686,8264794}.}
In \cite{7442079}, the authors maximized the successful computation probability of a single EU in a wirelessly powered MEC network by optimizing the time for energy harvesting (EH) and a binary computation offloading scheme, where the task is either executed locally or completely offloaded.
Extending the single EU scenario to multiple EUs, the authors proposed to maximize the weighted sum computation bits of all EUs by jointly optimizing the energy supply and binary computation offloading based on deep learning \cite{8771176} and  {\color{black}Convex optimization theory   \cite{8334188}}.
In  \cite{8234686} and \cite{8264794}, the authors proposed {\color{black}a partial computation offloading scheme}, where each task can be divided into
independent parts for offloading or local computing, and minimized the total energy consumption of the MEC server and the energy source by jointly optimizing the energy transmit beamforming, the EH time, and the partial computation offloading scheme, subject to the constraints on energy-causality and maximum computation latency without or with {\color{black}EUs' cooperation}, respectively.

In the above works, EUs offload task bits to the MEC server via {\color{black}active transmissions (ATs)} following the harvest-then-transmit (HTT) protocol. Due to the use of power consuming components, e.g., carrier  oscillator, AT may consume a large portion  of the harvested energy and leave very limited energy for local computation at EUs, leading to performance degradation.
On the contrary, the emerging backscatter communication (BackCom)  allows an EU to modulate its information on the incident signal and reflect it to the receiver and hence consumes much less energy than  AT \cite{CL}. Recently, BackCom has been considered in wirelessly powered MEC networks \cite{8849964,9014101}, where EUs can jointly utilize  BackCom and AT for task offloading. {\color{black}In \cite{8849964}, the authors studied a wireless sensor network with one hybrid
access point (HAP) and multiple EUs, and developed a price-based distributed time and workload allocation scheme to maximize a reward function of MEC offloading. In \cite{9014101}, the authors minimized the energy consumption of the HAP by jointly optimizing the time for BackCom and that for AT within a given time budget while ensuring the offloaded bits above a required amount.}
{\color{black}To the authors' best knowledge, the computation bits maximization problem has not been studied for backscatter assisted wirelessly powered MEC networks.}


{\color{black}In this paper, we study the weighted sum computation bits maximization problem for a backscatter assisted wirelessly powered MEC network comprising one PB, one MEC server and multiple EUs. Different from \cite{8849964,9014101}, we include both the computing frequencies and BackCom reflection coefficients of EUs as optimization variables, while considering BackCom circuit power consumption and a practical non-linear EH model, bringing new challenges for the resource allocation scheme design.}
Specifically, we formulate the problem as a joint optimization of the BackCom reflection coefficient, AT transmit power, computing frequency and execution time of each EU as well as the EUs' time allocation between BackCom and AT. {\color{black}Then, we transform the formulated non-convex problem into a convex one with the help of Convex optimization theory and the properties of the non-linear EH model, and derive closed-form expressions for the optimal reflection coefficient, transmit power, computing frequency and execution time of each EU.}
Performance of the proposed scheme in terms of weighted sum computation bits is evaluated through simulation in comparison with representative benchmark schemes{\footnote {\color{black} As the optimization objectives, system models,  power consumption and EH models used in \cite{8849964} and \cite{9014101} are different from those considered in this work, they are not included in the performance comparison.}}.

\vspace{-15pt}
\section{System Model}
As shown in Fig. 1, we consider a backscatter assisted wirelessly powered MEC network consisting of one PB, one MEC server and $K$ EUs, each with a rechargeable battery. In particular, each EU is {\color{black}equipped with} an EH module, backscatter circuit and {\color{black}an AT circuit} so that it can offload task bits to the MEC server via hybrid HTT and BackCom{\footnote{\color{black}
The main difference between our system model and the existing works on wireless powered MEC networks is the use of BackCom at each EU, which requires the EU being equipped with a backscatter circuit in addition to the EH module and the AT circuit. Since the backscatter circuit
could be a simple impedance matching circuit \cite{7820135}, it is possible to be equipped in each EU at a reasonably low cost. }}.
Assuming that the task bits of each task are bit-wise independent \cite{8234686,8264794}, we consider the partial offloading scheme. Similar to \cite{8334188,8234686,8264794}, we assume that each EU has separate computing circuit and offloading circuit so that each EU can perform local computation and task offloading simultaneously.
{\color{black}Following \cite{8334188,7542156}, we assume that each EU can adjust its computing frequency using the dynamic voltage scaling (DVS) technology. Note that although the realization of DVS may increase the hardware cost,
it can be justified by the resulting energy savings for EUs in the long term.}
Let $g_k$ $\left(k\in\{1,2,...,K\}\right)$ and $h_k$ be the channel gains of the PB-to-the $k$-th EU link and the MEC server-to-the $k$-th EU link, respectively. All the channels are  modeled as quasi-static fading.

\begin{figure}
  \centering
  \includegraphics[width=0.40\textwidth]{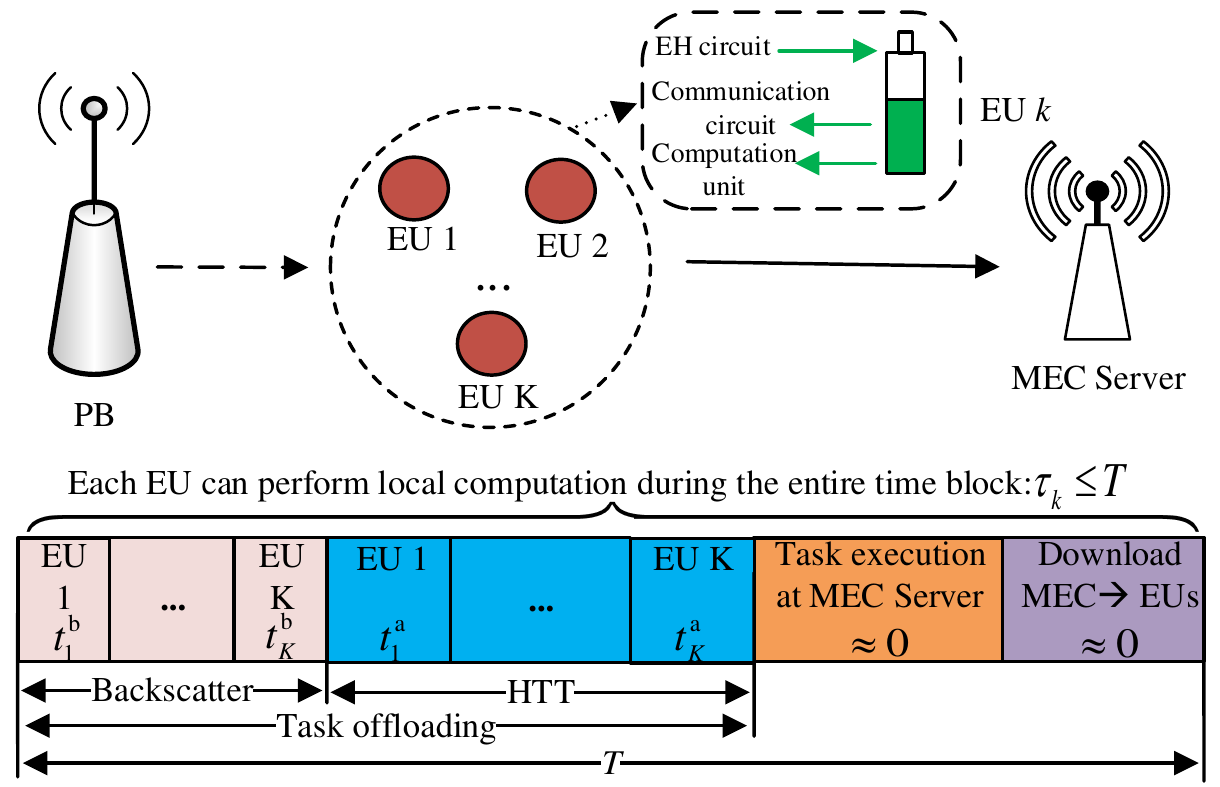}\\ \vspace{-5pt}
  \caption{System model and time allocation scheme.}\vspace{-10pt}
\end{figure}

Let $T$ denote the entire time block, which can be divided into four phases. In the first phase, the PB broadcasts the energy signals,  and all the EUs
take turns to  offload task bits to the MEC via BackCom, while the non-backscatter EUs work in the EH mode.
In the second phase, the PB stops broadcasting and all the EUs take turns to offload their task bits to the MEC server by  AT. The third phase is the task execution phase, where the MEC server executes all the received computation tasks. In the fourth phase, the MEC server will download the computation results to all the EUs. {\color{black} In this work, our designed scheme is mainly suitable for the applications where the MEC server is equipped with very high-performance CPUs and the computation results
of the MEC server are usually just a few bits, e.g., automatic manufacturing systems. Thus, the MEC server's computation time (the third phase) and the downloading time (the fourth phase) can be
ignored \cite{8234686,8264794}.}

Let $t^{\rm{b}}_k$ denote the time allocated to the $k$-th EU for backscattering in the first phase. {\color{black}During the sub-phase $t^{\rm{b}}_k$, the $k$-th EU will divide the received PB signal into two parts according to a changeable reflection
coefficient $\alpha_k$ $\left(0\leq\alpha_k\leq1\right)$ \cite{7820135}: one part is backscattered to the MEC server carrying some task bits, and the rest is fed into the EH circuit.}
For BackCom, we assume that successive interference cancellation (SIC) is performed at the MEC server to remove the interference caused by the PB-to-MEC server link{\footnote{{\color{black}In this work, the PB serves only as a RF power source, and hence the RF signal transmitted by the PB can be predetermined and known to the MEC server. Following the channel estimation procedures in \cite{CL}, the MEC server obtains all the instantaneous channel state information (CSI), and thus can remove the interference from the PB by performing the SIC and  determine the optimal resource allocation policy, which is then transmitted to the EUs and the PB.}}}.
{\color{black}Then, based on \cite{8802296}, the achievable offloading throughput of the $k$-th EU via BackCom in the first phase can be expressed as
$R^{\rm{b}}_{k}=t_k^{\rm{b}}B{\log _2}\left( {1 + \frac{{\xi {\alpha _k}{P_{\rm{t}}}{g_k}{h_k}}}{{B{\sigma ^2}}}} \right)$,
where  $\xi$ is the performance
gap between the BackCom and the AT \cite{8802296,7981380}, $B$, $P_{\rm{t}}$ and $\sigma^2$ denote the channel bandwidth, the PB's transmit power, and the thermal noise power spectral density, respectively.}

For EH, we employ a  practical non-linear EH model  \cite{7999248} to characterize the EH circuit. Thus, the harvested energy at the $k$-th EU during the sub-phase $t^{\rm{b}}_k$ is computed as
$E^{\rm{b}}_{k}=\left(\frac{{{c_k}\left( {1 - {\alpha _k}} \right){P_{\rm{t}}}{g_k} + {d_k}}}{{\left( {1 - {\alpha _k}} \right){P_{\rm{t}}}{g_k} + {v_k}}} - \frac{{{d_k}}}{{{v_k}}}\right)t^{\rm{b}}_k$,
where $c_k$, $d_k$ and $v_k$ are the parameters of the non-linear EH model at the $k$-th EU.
Accordingly, at the end of the first phase, the total harvested energy at the $k$-th EU is given by
$E^{\rm{total}}_{k}=E^{\rm{b}}_k+P^{\rm{h}}_k\left(\sum\nolimits_{i = 1}^K {t_i^{\rm{b}} - t_k^{\rm{b}}} \right)$,
where $P^{\rm{h}}_k=\frac{{{c_k}{P_{\rm{t}}}{g_k} + {d_k}}}{{{P_{\rm{t}}}{g_k} + {v_k}}} - \frac{{{d_k}}}{{{v_k}}}$.
Denote the transmit power and time for the $k$-th EU in the second phase by $p_k$ and $t^{\rm{a}}_k$, respectively. Then the  offloading throughput of the $k$-th EU via AT during $t^{\rm{a}}_k$ is given by
$R^{\rm{a}}_k=t_k^{\rm{a}}B{\log _2}\left( {1 + \frac{{{p_k}{h_k}}}{{B{\sigma ^2}}}} \right)$.
Then we obtain the total offloading throughput of the $k$-th EU as $R^{\rm{o}}_k=R^{\rm{b}}_k+R^{\rm{a}}_k$.

For local computing, let $f_k$ and $\tau_k$ $\left(0\leq \tau_k\leq T\right)$ be the local computing frequency and  execution time of the $k$-th EU, respectively. Based on \cite{8234686}, the computation bits and the computation energy consumption at the $k$-th EU can be calculated as $R^{\rm{e}}_k=\frac{\tau_k f_k}{C_{\rm{cpu}}}$, and $E^{\rm{e}}_k={\varepsilon _{k}}f_{k}^3\tau_k,$ respectively,
where $C_{\rm{cpu}}$ is the number of CPU cycles required for computing one bit and $\varepsilon _{k}$ is the effective capacitance coefficient of the processor's chip at the $k$-th EU.
\vspace{-10pt}
\section{Computation Bits Maximization}
\subsubsection{Problem Formulation}
We propose a scheme to maximize the weighted sum computation bits of all the EUs in each time block by jointly optimizing the EUs' BackCom time allocation $[t^{\rm{b}}_1,\cdots,t^{\rm{b}}_K]$ and reflection coefficients $[\alpha_1,\cdots,\alpha_K]$, AT transmit time $[t^{\rm{a}}_1,\cdots,t^{\rm{a}}_K]$ and power $[p_1,\cdots,p_K]$, and local computing frequencies $[f_1,\cdots,f_K]$ and execution time $[\tau_1,\cdots,\tau_K]$.
{\color{black}We assume a constant circuit power consumption rate for BackCom \cite{CL}. Let $P_{c,k}$ and $p_{c,k}$ denote the circuit power consumption for BackCom and for AT at the $k$-th EU, respectively. Then, the consumed energy for BackCom and for AT at the $k$-th EU can be computed as ${P_{c,k}}t_k^{\rm{b}}$ and $\left( {{p_k} + {p_{c,k}}} \right)t_k^{\rm{a}}$, respectively.
Let $w_k>0$ denote the weight of the $k$-th EU, which indicates the priority of the $k$-th EU in the weighted sum computation bits maximization problem. Note that these predefined weights can be used to customize the service provisioning for different EUs.} Then, the computation bits maximization problem is formulated as
\begin{align}\notag
\begin{array}{*{20}{l}}
{{{\bf{P}}_{\bf{0}}}:\;\mathop {\max }\limits_{{\bf{t^{\rm{b}}}},{\bm{\alpha}} ,{\bf{p}},{\bf{t^{\rm{a}}}},{\bf{f}},{\bm{\tau}} } \sum\nolimits_{k = 1}^K {{w_k}\left( {R_k^{\rm{o}} + R_k^{\rm{e}}} \right)} }\\
{\begin{array}{*{20}{l}}
{{\rm{s}}.{\rm{t}}.\;{\rm{C1}}:\;R_k^{\rm{o}} + R_k^{\rm{e}} \ge {L_{\min ,k}},\forall k,}\\
{\;\;\;\;\;\;{\rm{C}}2:{P_{c,k}}t_k^{\rm{b}} + \left( {{p_k} + {p_{c,k}}} \right)t_k^{\rm{a}} + {\varepsilon _k}f_k^3{\tau _k} \le E_k^{{\rm{total}}},\;\forall k,}
\end{array}}\\
{\begin{array}{*{20}{l}}
{\;\;\;\;\;\;{\rm{C}}3:\;\sum\nolimits_{k = 1}^K {\left( {t_k^{\rm{b}} + t_k^{\rm{a}}} \right)}  \le T,\;0 \le {\tau _k} \le T,\forall k,}\\
{\;\;\;\;\;\;{\rm{C}}4:\;0 \le {f_k} \le f_k^{\max },\;\forall k,}
\end{array}}\\
{\begin{array}{*{20}{l}}
{\;\;\;\;\;\;{\rm{C}}5:\;0 \leq {\alpha _k} \leq 1,\forall k,\;}\\
{\;\;\;\;\;\;{\rm{C}}6:\;t_k^{\rm{b}} \ge 0,\;t_k^{\rm{a}} \ge 0,{p_k} \ge 0,\forall k,}
\end{array}}
\end{array}
\end{align}
where ${\bf{t^{\rm{b}}}}=[t^{\rm{b}}_1,\cdots,t^{\rm{b}}_K]$, ${\bm{\alpha}}=[\alpha_1,\cdots,\alpha_K]$, ${\bf{p}}=[p_1,\cdots,p_K]$, ${\bf{t^{\rm{a}}}}=[t^{\rm{a}}_1,\cdots,t^{\rm{a}}_K]$, ${\bf{f}}=[f_1,\cdots,f_K]$, ${\bm{\tau}}=[\tau_1,\cdots,\tau_K]$,
$L_{\min ,k}$ denotes the minimum required computation bits for the $k$-th EU, 
and $f_k^{\rm{max}}$ denotes the maximum CPU frequency of the $k$-th EU.
In ${{\bf{P}}_{\bf{0}}}$, $\rm{C}1$ guarantees the minimum required computation task bits for each EU.
$\rm{C}2$ and $\rm{C}3$ are the energy-causality and the time allocation constraints.


\subsubsection{Solution}
Problem ${{\bf{P}}_{\bf{0}}}$ is non-convex due to the non-linear EH model and the coupling relationships between the optimization variables (i.e., $\alpha_k$ and $t^{\rm{b}}_k$, $f_k$ and $\tau_k$, etc.) in both the objective function and constraints, i.e., $\rm{C}1$ and $\rm{C}2$. {\color{black}Specifically, the
use of the non-linear EH model will make $\rm{C}2$ complicated and non-convex.}
To solve ${{\bf{P}}_{\bf{0}}}$, we provide the following lemma to obtain the optimal execution time for each EU.

{\textit{Lemma 1.}} Let $\tau_k^*$ $\left(k\in\{1,2,...,K\}\right)$ denote the optimal execution time of the $k$-th EU. Then the maximum weighted sum computation
bits of all the EUs can be achieved when each EU performs local computing throughout each time block, i.e., $\tau_k^*=T$.

\emph{Proof.} Please see Appendix A. \hfill {$\blacksquare $}

Substituting $\tau_k^*=T$ into ${{\bf{P}}_{\bf{0}}}$, we have
\begin{align}\notag
\begin{array}{*{20}{l}}
{{{\bf{P}}_1}:\;\mathop {\max }\limits_{{{\bf{t}}^{\rm{b}}},{\bm{\alpha}} ,{\bf{p}},{{\bf{t}}^{\rm{a}}},{\bf{f}}} \sum\nolimits_{k = 1}^K {{w_k}\left( {R_k^{\rm{o}} + \frac{{{f_k}T}}{{{C_{{\rm{cpu}}}}}}} \right)} }\\
{\begin{array}{*{20}{l}}
{{\rm{s}}.{\rm{t}}.\;{\rm{C1 - 1}}:\;R_k^{\rm{o}} + \frac{{{f_k}T}}{{{C_{{\rm{cpu}}}}}} \ge {L_{\min ,k}},\forall k,}\\
{\;\;\;\;\;\;{\rm{C}}2 - 1:{P_{c,k}}t_k^{\rm{b}} + \left( {{p_k} + {p_{c,k}}} \right)t_k^{\rm{a}} + {\varepsilon _k}f_k^3T \le E_k^{{\rm{total}}},\;\forall k,}
\end{array}}\\
{\begin{array}{*{20}{l}}
{\;\;\;\;\;\;{\rm{C}}3 - 1:\;\sum\nolimits_{k = 1}^K {\left( {t_k^{\rm{b}} + t_k^{\rm{a}}} \right)}  \le T,}\\
{\;\;\;\;\;\;{\rm{C}}4,{\rm{C}}5,{\rm{C6}}{\rm{.}}}
\end{array}}
\end{array}
\end{align}

Although ${{\bf{P}}_{\bf{1}}}$ is more tractable than ${{\bf{P}}_{\bf{0}}}$, it is still non-convex due to coupling relationships between $\alpha_k$ and $t^{\rm{b}}_k$, and between $p_k$ and $t^{\rm{a}}_k$.
To tackle this issue, we introduce the following
auxiliary variables: $x_k=\alpha_kt^{\rm{b}}_k$ and $P_k=p_kt^{\rm{a}}_k$ $\left(\forall k\in\{1,2,...,K\}\right)$ into ${{\bf{P}}_{\bf{1}}}$ and reformulate ${{\bf{P}}_{\bf{1}}}$ as
\begin{align}\notag
\begin{array}{*{20}{l}}
{{{\bf{P}}_2}:\;\mathop {\max }\limits_{{{\bf{t}}^{\rm{b}}},{\bf{x}},{\bf{P}},{{\bf{t}}^{\rm{a}}},{\bf{f}}} \sum\nolimits_{k = 1}^K {{w_k}\left( {C_k^{\rm{o}}({x_k},t_k^{\rm{b}},{P_k},t_k^{\rm{a}}) + \frac{{{f_k}T}}{{{C_{{\rm{cpu}}}}}}} \right)} }\\
{\begin{array}{*{20}{l}}
{{\rm{s}}.{\rm{t}}.\;{\rm{C1 - 2}}:\;C_k^{\rm{o}}({x_k},t_k^{\rm{b}},{P_k},t_k^{\rm{a}}) + \frac{{{f_k}T}}{{{C_{{\rm{cpu}}}}}} \ge {L_{\min ,k}},\forall k,}\\
\begin{array}{l}
\;\;\;\;{\rm{C}}2 - 2:{P_{c,k}}t_k^{\rm{b}} + {P_k} + {p_{c,k}}t_k^{\rm{a}} + {\varepsilon _k}f_k^3T \le \\
\;\;\;\;\;\;\;\;\;\;\;\;\;\;\;\;\;\;N_k^{\rm{b}}\left( {{x_k},t_k^{\rm{b}}} \right) + P_k^{\rm{h}}\left( {\sum\nolimits_{i = 1}^K {t_i^{\rm{b}} - t_k^{\rm{b}}} } \right),\;\forall k,
\end{array}
\end{array}}\\
{\begin{array}{*{20}{l}}
{\;\;\;\;\;\;{\rm{C}}3 - 1,{\rm{C}}4,{{\rm{C}}5 - 1:0 \leq {x_k} \leq t_k^{\rm{b}},\forall k,}}\\
{\begin{array}{*{20}{l}}
{\;\;\;\;{\rm{C6 - 1}}:t_k^{\rm{b}} \ge 0,\;t_k^{\rm{a}} \ge 0,{P_k} \ge 0,\forall k,}
\end{array}}
\end{array}}
\end{array}
\end{align}
where $C^{\rm{o}}_{k}(x_k,t^{\rm{b}}_k,P_k,t^{\rm{a}}_k)=t_k^{\rm{b}}B{{\log }_2}\left( {1 + \frac{{\xi {x_k}{P_{\rm{t}}}{g_k}{h_k}}}{{t_k^{\rm{b}}B{\sigma ^2}}}} \right) + t_k^{\rm{a}}B{{\log }_2}\left( {1 + \frac{{{P_k}{h_k}}}{{t_k^{\rm{a}}B{\sigma ^2}}}} \right)$ and $N_k^{\rm{b}}\left( {{x_k},t_k^{\rm{b}}} \right) = \left( {\frac{{{c_k}\left( {1 - \frac{{{x_k}}}{{t_k^{\rm{b}}}}} \right){P_{\rm{t}}}{g_k} + {d_k}}}{{\left( {1 - \frac{{{x_k}}}{{t_k^{\rm{b}}}}} \right){P_{\rm{t}}}{g_k} + {v_k}}} - \frac{{{d_k}}}{{{v_k}}}} \right)t_k^{\rm{b}}$.

{\color{black}It is difficult to tell whether ${{\bf{P}}_{\bf{2}}}$ is convex or not due to the use of the non-linear EH model. In what follows, Proposition 1 is provided to tackle ${{\bf{P}}_{\bf{2}}}$.

\textbf{Proposition 1.} The optimization problem ${{\bf{P}}_{\bf{2}}}$ is proved to be convex with the help of Convex optimization theory and the properties of the non-linear EH model.

{\color{black}\emph{Proof.} As shown in ${{\bf{P}}_{\bf{2}}}$, $\rm{C3-1}$, $\rm{C4}$, $\rm{C5-1}$ and $\rm{C6-1}$ are linear constraints and whether ${{\bf{P}}_{\bf{2}}}$ is convex or not depends on the objective function and constraints $\rm{C1-2}$ and $\rm{C2-2}$. Specifically, as for the objective function and $\rm{C1-2}$, $\frac{f_kT}{C_{\rm{cpu}}}$ is a linear function with respect to $f_k$ and the function $F_0(x,y)=x\log_2\left(1+\frac{y}{x}\right)$ must be concave to ensure the concave objective function and the convex $\rm{C1-2}$. Note that $F_0(x,y)$ is the perspective of $\log_2\left(1+y\right)$ that is a concave function. Since the perspective operation preserves convexity \cite{Cvex}, $F_0(x,y)$ is a concave function in regard to $x$ and $y$.

As for $\rm{C2-2}$, the left side ${P_{c,k}}t_k^{\rm{b}} + {P_k} + {p_{c,k}}t_k^{\rm{a}} + {\varepsilon _k}f_k^3T$ is a linear function regarding $t_k^{\rm{b}}$, $P_k$ and $t_k^{\rm{a}}$. Since $f_k\geq 0$, ${P_{c,k}}t_k^{\rm{b}} + {P_k} + {p_{c,k}}t_k^{\rm{a}} + {\varepsilon _k}f_k^3T$ is a convex function with respect to $f_k$.
The right side of $\rm{C2-2}$ is $N^{\rm{b}}_k(x_k,t_k^{\rm{b}})=\left( {\frac{{{c_k}\left( {1 - \frac{{{x_k}}}{{t_k^{\rm{b}}}}} \right){P_{\rm{t}}}{g_k} + {d_k}}}{{\left( {1 - \frac{{{x_k}}}{{t_k^{\rm{b}}}}} \right){P_{\rm{t}}}{g_k} + {v_k}}} - \frac{{{d_k}}}{{{v_k}}}} \right)t_k^{\rm{b}}$. If $N^{\rm{b}}_k(x_k,t_k^{\rm{b}})$ is a concave function regarding $x_k$ and $t_k^{\rm{b}}$, then $\rm{C2-2}$ is a convex constraint.
Likewise, based on the perspective function, we can draw that the convexity of $N^{\rm{b}}_k(x_k,t_k^{\rm{b}})$ is same as the function $F_k(x_k)={\frac{{{c_k}\left( {1 - x_k} \right){P_{\rm{t}}}{g_k} + {d_k}}}{{\left( {1 - x_k} \right){P_{\rm{t}}}{g_k} + {v_k}}} - \frac{{{d_k}}}{{{v_k}}}} $ $(0<x_k<1)$.
By taking the second-order derivative of $F_k(x_k)$ with respect to $x_k$, we have
$\frac{{{\partial ^2}{F_k}}}{{\partial {x^2_k}}} = \frac{{2{{\left( {{P_{\rm{t}}}{g_k}} \right)}^2}\left( {{d_k} - {c_k}{v_k}} \right)}}{{{{\left( {\left( {1 - {x_k}} \right){P_{\rm{t}}}{g_k} + {v_k}} \right)}^3}}}.$

From the expression of $\frac{{{\partial ^2}{F_k}}}{{\partial {x^2_k}}}$, we can see that the function $F_k(x_k)$ is concave or not depends on the signs of ${{d_k} - {c_k}{v_k}}$ and ${\left( {1 - {x_k}} \right){P_{\rm{t}}}{g_k} + {v_k}}$ within $0<x_k<1$. In the following part, we will prove that both ${{d_k} - {c_k}{v_k}}\leq0$ and $v_k\geq0$ always hold by using the properties of the non-linear EH model.
Firstly, the harvested power increases with the increasing of the input power and then converges to the maximum value when the input power is large enough. This means that the first-order derivative of $F_k(x_k)$ in regard to $\left({1 - {x_k}} \right){P_{\rm{t}}}{g_k}$ is not less than 0, given by
$\frac{{\partial {F_k}}}{{\partial \left( {1 - {x_k}} \right){P_{\rm{t}}}{g_k}}} = \frac{{{c_k}{v_k}-{d_k}}}{{{{\left( {\left( {1 - {x_k}} \right){P_{\rm{t}}}{g_k} + {v_k}} \right)}^2}}}\geq 0.$
Based on the expression of $\frac{{\partial {F_k}}}{{\partial \left( {1 - {x_k}} \right){P_{\rm{t}}}{g_k}}}$, we have ${{d_k} - {c_k}{v_k}}\leq0$.
Besides, the maximum harvestable power is not less than 0, i.e., $\mathop {\lim }\limits_{\left( {1 - {x_k}} \right){P_{\rm{t}}}{g_k} \to \infty } {F_k} = {c_k} - \frac{{{d_k}}}{{{v_k}}} = \frac{{{c_k}{v_k} - {d_k}}}{{{v_k}}}\geq0$. Thus, $v_k\geq0$ can be obtained. Combining $v_k\geq0$ and ${{d_k} - {c_k}{v_k}}\leq0$, $\frac{{{\partial ^2}{F_k}}}{{\partial {x^2_k}}}\leq0$ holds and $F_k(x_k)$ is a concave function with respect to $x_k$.
Accordingly, ${{\bf{P}}_{\bf{2}}}$ is proved to be convex and can be solved by using existing convex methods (i.e., interior point method, Lagrange duality, etc) efficiently. The proof is completed. \hfill {$\blacksquare $}}

{\color{black}Assuming that the interior point method is used to obtain the optimal solution to ${{\bf{P}}_{\bf{2}}}$, the computational complexity for solving ${{\bf{P}}_{\bf{2}}}$ is given by $O\left( {\sqrt {{m_1}} \log \left( {{m_1}} \right)} \right)$ \cite{Cvex}, where $m_1$ denotes the number of inequality constraints of ${{\bf{P}}_{\bf{2}}}$.}
By means of the Lagrange duality method, we provide the following theorem for obtaining closed-form expressions of the optimal reflection coefficient, transmit power and computing frequency of each EU.

\textbf{Theorem 1.} Given the non-negative Lagrange multipliers, i.e., $\bm{\theta }=\left( {{\theta _1},{\theta _2}, \cdots ,{\theta _K}} \right)$, $\bm{\mu }=\left( {{\mu _1},{\mu _2}, \cdots ,{\mu _K}} \right)$, $\bm{\varphi }=\left( {{\varphi _1},{\varphi _2}, \cdots ,{\varphi _K}} \right)$ and $\bm{\vartheta }=\left( {{\vartheta _0},{\vartheta _1}, \cdots ,{\vartheta _K}} \right)$, parts of the optimal solutions to ${{\bf{P}}_{\bf{2}}}$ can be obtained as follows,
\begin{small}
 \begin{align}\label{11}
&f_k^* = {\left[ {\sqrt {\frac{{\left( {{w_k} + {\theta _k}} \right)T - {\varphi _k}{C_{{\rm{cpu}}}}}}{{3{\mu _k}{\varepsilon _k}T{C_{{\rm{cpu}}}}}}} } \right]^ + },\\
&p_k^*={\left[ {\frac{{\left( {{w_k} + {\theta _k}} \right)B}}{{{\mu _k}\ln2}} - \frac{{B{\sigma ^2}}}{{{h_k}}}} \right]^ + },\\
&\alpha _k^* = \left\{ {\begin{array}{*{20}{c}}
{{{\left[ {\frac{{{B_k} - \sqrt {B_k^2 - 4{A_k}{D_k}} }}{{2{A_k}}}} \right]}^ + },{\vartheta _k} = 0}\\
{1,{\vartheta _k} > 0}
\end{array}} \right.,
\end{align}
\end{small}where ${\left[ x \right]^ + } = \max \left\{ {x,0} \right\}$, ${A_k} = \frac{{\left( {{w_k} + {\theta _k}} \right)B\xi P_{\rm{t}}^3g_k^3{h_k}}}{{\ln2}}$, $B_k = 2{A_k} + \frac{{2\left( {{w_k} + {\theta _k}} \right)B\xi P_{\rm{t}}^2g_k^2{h_k}{v_k}}}{{\ln2}} + \xi {\mu _k}\left( {{c_k}{v_k} - {d_k}} \right){P_{\rm{t}}}{g_k}{h_k}$, and ${D_k} = {A_k} + \frac{{\left( {{w_k} + {\theta _k}} \right)B\xi {P_{\rm{t}}}{g_k}{h_k}v_k^2}}{{\ln2}} + \frac{{2\left( {{w_k} + {\theta _k}} \right)B\xi P_{\rm{t}}^2g_k^2{h_k}{v_k}}}{{\ln2}} - {\mu _k}\left( {{c_k}{v_k} - {d_k}} \right)B{\sigma ^2}$.

\emph{Proof.} Please see Appendix B. \hfill {$\blacksquare $}

\emph{Remark 1.} From (1), we can see that if there are task bits to be locally computed, the optimal computing frequency of each EU may increase with the increasing weight of each EU.
From (2), it can be observed that each EU chooses to offload task bits to the MEC server during the second phase only when the channel gain between the MEC server and the EU is good enough, i.e., ${{h_k} > \frac{{{\sigma ^2}{\mu _k}\ln 2}}{{{w_k} + {\theta _k}}}}$ must hold to ensure a non-zero transmit power. Based on (1) and (2), we find that $\mu_k>0$ always holds. Combining its associated complementary slackness condition, we have ${P_{c,k}}t_k^{{\rm{b*}}} + P_k^* + {p_{c,k}}t_k^{{\rm{a*}}} + {\varepsilon _k}{\left( {f_k^*} \right)^3}T = N_k^{\rm{b}}\left( {x_k^*,t_k^{{\rm{b*}}}} \right) + P_k^{\rm{h}}\left( {\sum\nolimits_{i = 1}^K {t_i^{{\rm{b*}}} - t_k^{{\rm{b*}}}} } \right)$. This means that each EU consumes all the harvested energy for maximizing the weighted sum computation bits of all the EUs. From (3), we can see that when $D_k>0$, i.e., ${h_k} > \frac{{{\mu _k}\left( {{c_k}{v_k} - {d_k}} \right){\sigma ^2}\ln 2}}{{\left( {{w_k} + {\theta _k}} \right)\left( {P_{\rm{t}}^3g_k^3 + {P_{\rm{t}}}{g_k}v_k^2 + 2P_{\rm{t}}^2g_k^2{v_k}} \right)\xi }}$, the $k$-th EU performs BackCom in the first phase, which will increase  its computation bits.


\vspace{-20pt}
\section{Numerical Results}
In this section, we evaluate the performance of the proposed scheme via computer simulations. Unless otherwise specified, the basic simulation parameters are given as: $T=1$s, $B=100$kHz, $\sigma^2=-120$dBm/Hz, $C_{\rm{cpu}}=1000$Cycles/bit, $K=4$, $P_{\rm{t}}=3$W, $\xi=-15$dB, $w_1=w_2=w_3=w_4=1$, $P_{c,1}=P_{c,2}=P_{c,3}=P_{c,4}=100\mu$W, $p_{c,1}=p_{c,2}=p_{c,3}=p_{c,4}=1$mW, $\varepsilon_1=\varepsilon_2=\varepsilon_3=\varepsilon_4=10^{-26}$, $f^{\max}_1=f^{\max}_2=f^{\max}_3=f^{\max}_4=5\times10^8$Hz and $L_{\min,1}=L_{\min,2}=L_{\min,3}=L_{\min,4}=L_{\min}=20$kbits.
We consider the standard power loss propagation model for modeling the channel gains of the PB-to-the $k$-th EU link and the $k$-th EU-to-the MEC server link. Specifically, $g_k=g'_kd_{0k}^{-\beta}$ and $h_k=h'_kd_{1k}^{-\beta}$, where $g'_k$ and $h'_k$ denote the corresponding small-scale fading, $d_{0k}$ and $d_{1k}$ are the distances from the $k$-th EU to the PB and the MEC server, and $\beta$ is the path loss exponent.
We set $\beta=3$, $d_{01}=12$m, $d_{02}=10$m, $d_{03}=15$m, $d_{04}=13$m, $d_{11}=30$m, $d_{12}=35$m, $d_{13}=20$m and $d_{14}=25$m.
According to \cite{7999248}, the specific parameters of the used non-linear EH model are set as: $c_1=c_2=c_3=c_4=2.463$, $d_1=d_2=d_3=d_4=1.635$ and $v_1=v_2=v_3=v_4=0.826$.

In order to illustrate the superiority of the proposed scheme, we consider the following four representative benchmark schemes:
1) Complete offloading: all the EUs offload their whole task bits to the MEC server via hybrid HTT and BackCom;
2) Fully local computing: all the EUs perform computation locally;
3) Pure backscatter mode: each EU can offload part of task bits to the MEC server by BackCom and perform local computing at the same time;
4) Pure HTT mode: each EU uses its harvested energy to transmit part of task bits to the MEC server following the HTT protocol and perform local computing simultaneously. Note that the above four schemes are optimized under the same constraints as ${{\bf{P}}_{\bf{0}}}$ and can be regarded as special cases for the proposed scheme.

\begin{figure*}
\begin{align}\notag
{\cal L} &= \!\!\sum\limits_{k = 1}^K \!{\left[ \!{\left( {{w_k} \!+\! {\theta _k}} \right)\!\left( \!\! {t_k^{\rm{b}}B{{\log }_2}\left(\! {1 \!+ \frac{{\xi {x_k}{P_{\rm{t}}}{g_k}{h_k}}}{{t_k^{\rm{b}}B{\sigma ^2}}}} \!\right)\! +\! t_k^{\rm{a}}B{{\log }_2}\!\left(\! {1 + \frac{{{P_k}{h_k}}}{{t_k^{\rm{a}}B{\sigma ^2}}}} \right)\! +\! \frac{{{f_k}T}}{{{C_{{\rm{cpu}}}}}}} \!\right)\! -\! {\theta _k}{L_{\min ,k}}} \right]}\!\!+\!\! \sum\limits_{k = 1}^K \!{{\varphi _k}}\! \left(\! {f_k^{\max }\!\!\!- \!{f_k}} \right)\\
& +\!\! \sum\limits_{k = 1}^K \!{{\mu _k}\!\left(\! {N_k^{\rm{b}}\!\left( {{x_k},t_k^{\rm{b}}} \right) \!+\! P_k^{\rm{h}}\!\left(\! {\sum\limits_{i = 1}^K {t_i^{\rm{b}} - t_k^{\rm{b}}} } \right)\! -\! {P_{c,k}}t_k^{\rm{b}} - {P_k} \!-\! {p_{c,k}}t_k^{\rm{a}} \!-\! {\varepsilon _k}f_k^3T}\! \right)}\! +\! {\vartheta _0}\!\left(\!\! {T \!- \!\sum\limits_{k = 1}^K\!\! {\left( {t_k^{\rm{b}} \!+ \!t_k^{\rm{a}}} \right)} } \!\!\right)+\!\! \sum\limits_{k = 1}^K\! {{\vartheta _k}\!\left( {t_k^{\rm{b}} \!-\! {x_k}} \right)}.\tag{B.1}
\end{align}
\hrulefill
\vspace{-10pt}
\end{figure*}

Fig. 2 shows the weighted sum computation bits versus the minimum required computation bits of each EU $L_{\min}$, where the proposed scheme and the above four schemes are considered.
It can be observed that the weighted sum computation bits under all the schemes will decrease with the increasing of $L_{\min}$ since for a larger $L_{\min}$, more resources will be allocated to the EUs with small computation bits, leading to a reduction of the weighted sum computation bits. By comparisons, we can also see that the proposed scheme can achieve the highest weighted sum computation bits as the proposed scheme provides more flexibility to utilize the resource efficiently.
Besides, the weighted sum computation bits under the proposed scheme are higher than those under the pure backscatter mode and the pure HTT mode, which illustrates the advantages of the combination of BackCom and the HTT protocol.

\begin{figure}
  \centering
  \includegraphics[width=0.35\textwidth]{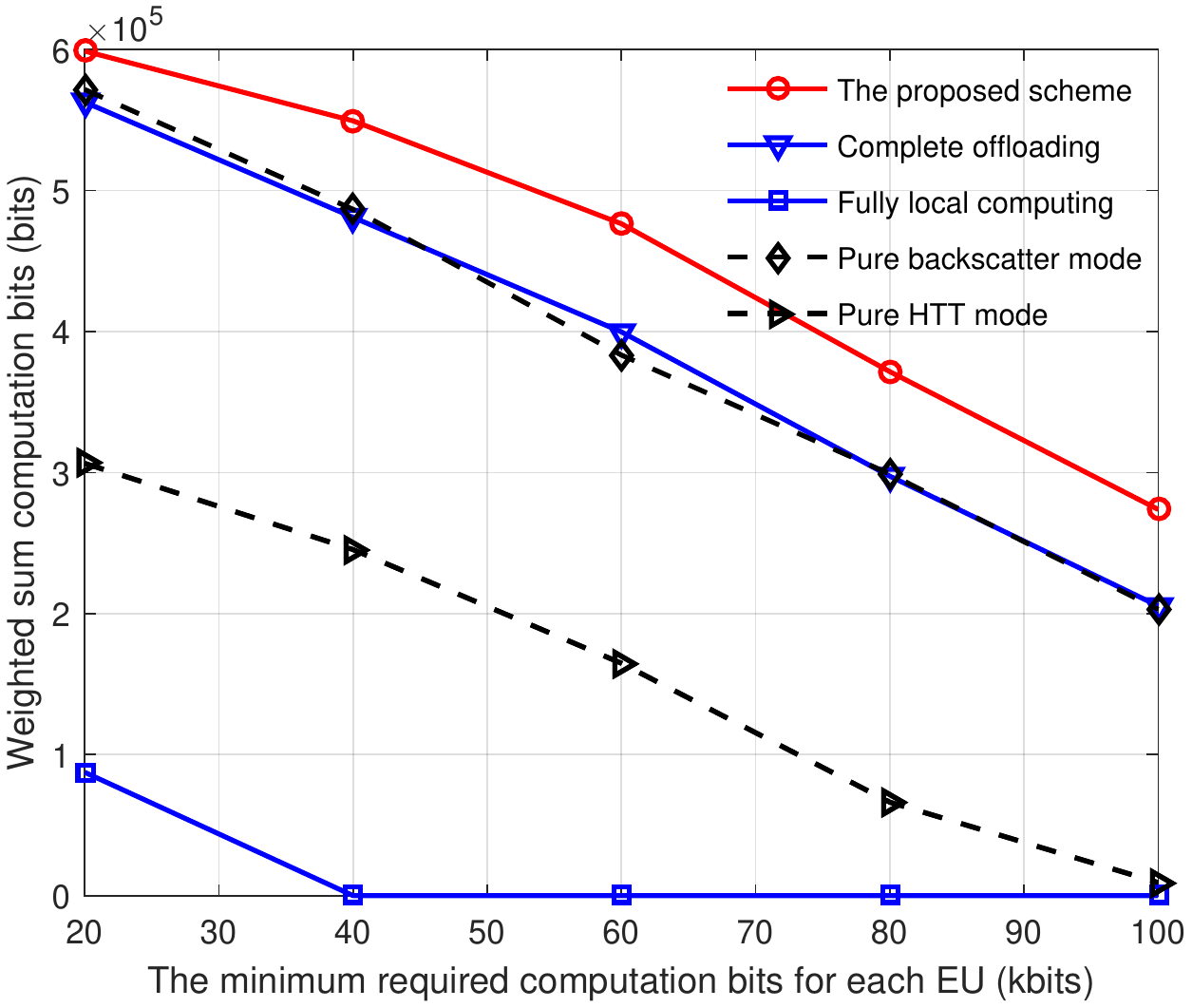}\\ \vspace{-3pt}
  \centering \caption{Weighted sum computation bits versus $L_{\min}$.}\vspace{-15pt}
\end{figure}

\vspace{-10pt}
\section{Conclusions}
In this paper, we have proposed a scheme to maximize the weighted sum computation bits in a backscatter assisted wirelessly powered MEC network, where a practical non-linear EH model and a flexible partial offloading scheme are considered for multiple EUs. {\color{black}Specifically, we formulated a weighted sum computation bits maximization problem by jointly optimizing the BackCom reflection coefficient and time, AT transmit power and time, local computing frequency and execution time of each EU, and transformed it into a convex optimization problem by introducing a series of auxiliary variables and using the properties of the non-linear EH model.}
Furthermore, we obtained the closed-form expressions for parts of the optimal solutions and provided insights into the maximization of weighted sum computation bits. Simulation results have confirmed that the proposed scheme outperforms the other schemes in terms of weighted sum computation bits.

\vspace{-5pt}
\section*{Appendix A}
{\color{black}When ${\bf{t^{\rm{b}}}},{\bm{\alpha}} ,{\bf{p}},{\bf{t^{\rm{a}}}}$ and $\{f_i,\tau_i\}_{i=\{1,2,\ldots,K\}\setminus k}$ are fixed, we jointly optimize $f_k$ and $\tau_k$ ($\forall k\in\{1,2,\ldots,K\}$) to maximize the weighted sum computation bits.
In the following, by means of contradiction, we will prove that the maximum weighted sum computation bits are achieved when $\tau_k^*=T$.
Specifically, let $f_k^*$ and $\tau_k^*$ $\left(\forall k\in\{1,2,...,K\}\right)$ denote the optimal computing frequency and execution time for the $k$-th EU, respectively.
Assume that $\tau_k^*<T$ and $\{f_k^*,\tau_k^*\}$ satisfies all the constraints of ${{\bf{P}}_{\bf{0}}}$ for given ${\bf{t^{\rm{b}}}},{\bm{\alpha}} ,{\bf{p}},{\bf{t^{\rm{a}}}}$ and $\{f_i,\tau_i\}_{i=\{1,2,\ldots,K\}\setminus k}$.
Then we construct another solution satisfying $\tau_k^+=T$ and $\tau_k^+(f_k^+)^3=\tau_k^*(f_k^*)^3$. Based on $\tau_k^+=T>\tau_k^*$, we can obtain $f_k^+<f_k^*$.
Thus, the constructed solution also satisfies the constraints of ${{\bf{P}}_{\bf{0}}}$.
Since $\tau_k^+f_k^+(f_k^+)^2=\tau_k^*f_k^*(f_k^*)^2$ and $f_k^+<f_k^*$, we have $\tau_k^+f_k^+>\tau_k^*f_k^*$. That is, the constructed solution can achieve a higher $R_k^{\rm{e}}$, leading to higher weighted sum computation bits.
This contradicts the above assumption that $\tau_k^*<T$.
Thus, Lemma 1 is proven.}

\vspace{-5pt}
\section*{Appendix B}
Let $\bm{\theta }=\left( {{\theta _1},{\theta _2}, \cdots ,{\theta _K}} \right)$, $\bm{\mu }=\left( {{\mu _1},{\mu _2}, \cdots ,{\mu _K}} \right)$, $\bm{\varphi }=\left( {{\varphi _1},{\varphi _2}, \cdots ,{\varphi _K}} \right)$ and $\bm{\vartheta }=\left( {{\vartheta _0},{\vartheta _1}, \cdots ,{\vartheta _K}} \right)$
denote the non-negative Lagrange multipliers with respect to all the constraints. Then the Lagrangian function of ${\bf{P}}_{\bf{2}}$ is given by (B.1), as shown at the top of the this page.
By taking the partial derivative of ${\cal L}$ with respect to $f_k$, $x_k$ and $P_k$, we have
\begin{align}
&\frac{{\partial {\cal L}}}{{\partial {f_k}}} = \frac{{\left( {{w_k} + {\theta _k}} \right)T}}{{{C_{{\rm{cpu}}}}}} - 3{\mu _k}{\varepsilon _k}Tf_k^2 - {\varphi _k},\tag{B.2}\\
&\frac{{\partial \mathcal{L}}}{{\partial {P_k}}} =\frac{{\left( {{w_k} + {\theta _k}} \right)t_k^{\rm{a}}B{h_k}}}{{\left( {t_k^{\rm{a}}B{\sigma ^2} + {P_k}{h_k}} \right)\ln2}} - {\mu _k},\tag{B.3}\\
&\frac{{\partial \mathcal{L}}}{{\partial {x_{k}}}}\!= \!\!\frac{{\left( {{w_k}\!\! + \!{\theta _k}} \right)\!t_k^{\rm{b}}B\xi {P_{\rm{t}}}{g_k}{h_k}}}{{\left( {t_k^{\rm{b}}B{\sigma ^2} \!\!+ \!\xi {x_k}{P_{\rm{t}}}{g_k}{h_k}} \!\right)\!\ln2}} \!-\! \frac{{{\mu _k}\!\left( {{c_k}{v_k} \!-\! {d_k}} \right)\!{{\left( {t_k^{\rm{b}}} \right)}^2}}}{{{{\left( \!{\left( \!{t_k^{\rm{b}}\! \!-\! {x_k}} \!\right)\!{P_{\rm{t}}}{g_k} \!\!+\! {v_k}t_k^{\rm{b}}} \!\right)}^2}}}\!-\!\! {\vartheta _k}.\tag{B.4}
\end{align}

By letting $\frac{{\partial {\cal L}}}{{\partial {f_k}}}=0$, we can compute the optimal CPU frequency of the $k$-th EU as
$f_k^* = {\left[ {\sqrt {\frac{{\left( {{w_k} + {\theta _k}} \right)T - {\varphi _k}{C_{{\rm{cpu}}}}}}{{3{\mu _k}{\varepsilon _k}T{C_{{\rm{cpu}}}}}}} } \right]^ + }$, 
where ${\left[ x \right]^ + } = \max \left\{ {x,0} \right\}$.
Then by letting $\frac{{\partial {\cal L}}}{{\partial {P_k}}}=0$ and $p_k=\frac{P_k}{t^{\rm{a}}_k}$, the optimal transmit power of the $k$-th EU during the second phase can be computed as
$p_k^*={\left[ {\frac{{\left( {{w_k} + {\theta _k}} \right)B}}{{{\mu _k}\ln2}} - \frac{{B{\sigma ^2}}}{{{h_k}}}} \right]^ + }$.

For (B.4), if $x_k=t^{\rm{b}}_k$, then the optimal reflection coefficient of the $k$-th EU is $\alpha^*_k=1$.
If $x_k<t^{\rm{b}}_k$ holds, ${\vartheta _k}=0$ must be satisfied based on the Karush-Kuhn-Tucker (KKT) conditions. In this case, the optimal reflection coefficient of the $k$-th EU
should satisfy the following equation, i.e.,
${A_k}{\left( {\alpha _k^*} \right)^2} - {B_k}\alpha _k^* + {D_k} = 0$, 
where ${A_k} = \frac{{\left( {{w_k} + {\theta _k}} \right)B\xi P_{\rm{t}}^3g_k^3{h_k}}}{{\ln2}}$, $B_k = 2{A_k} + \frac{{2\left( {{w_k} + {\theta _k}} \right)B\xi P_{\rm{t}}^2g_k^2{h_k}{v_k}}}{{\ln2}} + \xi {\mu _k}\left( {{c_k}{v_k} - {d_k}} \right){P_{\rm{t}}}{g_k}{h_k}$ and ${D_k} = {A_k} + \frac{{\left( {{w_k} + {\theta _k}} \right)B\xi {P_{\rm{t}}}{g_k}{h_k}v_k^2}}{{\ln2}} + \frac{{2\left( {{w_k} + {\theta _k}} \right)B\xi P_{\rm{t}}^2g_k^2{h_k}{v_k}}}{{\ln2}} - {\mu _k}\left( {{c_k}{v_k} - {d_k}} \right)B{\sigma ^2}$. Since $0\leq\alpha _k^*\leq1$, Thus, $\alpha _k^*$ under this case is determined by ${\left[ {\frac{{{B_k} - \sqrt {B_k^2 - 4{A_k}{D_k}} }}{{2{A_k}}}} \right]^ + }$ as ${\frac{{{B_k} + \sqrt {B_k^2 - 4{A_k}{D_k}} }}{{2{A_k}}}}>1$.
{\color{black}Although it is difficult to obtain closed-form expressions for the optimal $t_k^{\rm{b}}$ and $t_k^{\rm{a}}$, since the Lagrangian function $\cal L$ is a linear function of both $t_k^{\rm{b}}$ and $t_k^{\rm{a}}$ based on the obtained $\frac{{\partial {\cal L}}}{{\partial {t_k^{\rm{b}}}}}$ and $\frac{{\partial {\cal L}}}{{\partial {t_k^{\rm{a}}}}}$, standard linear optimization tools, such as the simplex method, can be used to obtain the optimal values of ${{\bf{t}}^{\rm{b}}}$ and ${{\bf{t}}^{\rm{a}}}$ efficiently.}
\vspace{-10pt}
\bibliographystyle{IEEEtran}
\bibliography{ref}
\end{document}